\documentclass[11pt]{article}
\usepackage{putex}
\usepackage[colorinlistoftodos]{todonotes}
\usepackage{tcolorbox}
\usepackage{amsfonts}
\usepackage{amssymb}
\usepackage{amsmath}
\usepackage[bookmarks = false, colorlinks = true, linkcolor = blue, citecolor = purple]{hyperref}
\usepackage{ulem}
\usepackage{tipa}
\usepackage{graphics,graphicx}
\usepackage{setspace}
\numberwithin{equation}{section}

\newcommand{\HH}{\mathcal{H}}

\newcommand{\Tr}{\mbox{Tr}}

\newcommand{\OO}{\mathcal{O}}

\newcommand\be{\begin{equation}}
\newcommand\ba{\begin{eqnarray}}
\newcommand\ee{\end{equation}}
\newcommand\ea{\end{eqnarray}}
\newcommand\bone{{\bf 1}}
\definecolor{lred}{rgb}{0.8,0,0}
\definecolor{lurple}{rgb}{0.8,0.0,0.6}
\definecolor{purple}{rgb}{0.5,0.0,0.3}
\definecolor{huh}{rgb}{0.0,0.6,0.8}
\definecolor{lorange}{rgb}{1,0.5,0}
\definecolor{orange}{rgb}{0.8,0.4,0}
\definecolor{dorange}{rgb}{0.6,0.2,0}
\definecolor{light-gray}{gray}{0.5}

\begin{document}

\title{Entanglement entropy in Jackiw-Teitelboim gravity 
\\ with matter}
\authors{Jennifer Lin\footnote{e-mail: \texttt{jylin04@gmail.com}}}
\institution{FHI}{{Future of Humanity Institute, University of Oxford, Oxford, OX2 0DJ, UK}}
\abstract{\noindent 
In this paper, I study the entanglement entropy in Hartle-Hawking states of JT gravity set up by a Euclidean path integral with an operator inserted somewhere along the Euclidean boundary. I show that the entanglement entropy between the dual SYK models can be written to leading order in $1/G_N$ in a way that resembles a formula for entanglement entropy in a compact lattice gauge theory, 
	but where the link that we appear to compute the entanglement across depends on where we put the operator along the Euclidean boundary. I then 
	describe a toy model 
	where we can write down an explicit family of density matrices whose von Neumann entropies exhibit such a phase transition as we tune an interpolating parameter. This behavior relies on a certain superselection sector structure and I comment on the possibility that the gravitational 
	result comes from a similar emergent superselection sector structure in SYK/JT holography. The main implication is that {each} copy of the SYK model should then contain Hilbert space sectors labeled by multiple $SL(2,R)$ representations whose entanglement gives rise to space.
}

\maketitle
\newpage

\section{Introduction} \label{s1}

The study of quantum entropy in gravitational systems has been surprisingly 
productive in the last fifteen years. This program was initiated by Ryu and Takayanagi \cite{Ryu:2006bv}, \cite{Ryu:2006ef} who taught us that in states of holographic CFTs with Einstein gravity duals, the fine-grained entropy of a region $A$ in the CFT is given to leading order in $1/G_N$ by the area of the minimal-area extremal surface $\gamma_A$ homologous to it in the bulk:
\be\label{rt}
S(A) = \frac{\mbox{Area}(\gamma_A)}{4G_N}\,.
\ee
This equation is of course suggestive of a deep connection between quantum entanglement and the Bekenstein-Hawking entropy of black holes.

The Ryu-Takayanagi formula has since been extended in many directions, to time-dependent situations \cite{Hubeny:2007xt}, $1/G_N$ corrections \cite{Faulkner:2013ana}, etc.; see \cite{Rangamani:2016dms} for a review. 
Most recently it was extended to the ``Island rule" \cite{Almheiri:2019hni} for computing the fine-grained entropy of a non-gravitational system $\mathsf{R}$ coupled to gravity:
\be\label{islandrule}
S(\mathsf{R}) =\min\left\{\mbox{ext}_{I} \left[\frac{\mbox{Area}(\partial I)}{4G_N} + S_{semi-cl}(\mathsf{R}\cup I)\right]\right\}\,,
\ee
where $I$ is some number of bulk regions (``islands") bounded by extremal surfaces, and $S_{semi-cl}(\mathsf{R}\cup I)$ is the fine-grained entropy of the combined quantum state of $\mathsf{R}$ and $I$ in the bulk semiclassical approximation. Surprisingly, \eqref{islandrule} allows us to compute the correct Page curve for the Hawking radiation in toy models of evaporating black holes \cite{Almheiri:2019hni}, \cite{Almheiri:2019psf}, \cite{Penington:2019npb}, \cite{Almheiri:2019qdq}, \cite{Penington:2019kki} (see \cite{Almheiri:2020cfm} for a review). 
However, we lack a microscopic understanding of eqs. \eqref{rt} and \eqref{islandrule}. 
We would like to understand when and why these 
formulas are true.

More precisely, we know how to derive \eqref{rt} and \eqref{islandrule} from a Euclidean path integral approach \cite{Almheiri:2019qdq}, \cite{Penington:2019kki}, \cite{Lewkowycz:2013nqa}, but we don't know what the gravitational path integral is computing from a canonical point of view. In other words, we don't know how to write down an exact family of density matrices whose von Neumann entropies in some parametric limit would give rise to these formulas.

In this paper, I will attempt to make progress on this problem by developing an analogy between emergent compact gauge theories and the holographic duality between the SYK model and $1+1d$ Jackiw-Teitelboim (JT) gravity. Namely, specializing to states in JT gravity that can be set up by a Euclidean path integral, I will (1) rewrite the entanglement entropy between the dual SYK models in an unusual way; (2) identify situations in (a particular extension of) a compact lattice gauge theory where we can write down an exact density matrix whose von Neumann entropy takes the same functional form; (3) conjecture that the compact gauge theory and gravity formulas arise due to similar underlying physics; 
	and (4) see what consequences this has for how (SYK/JT) holography might work. This strategy is illustrated in Figure \ref{fig1}. 

By applying this strategy to vacuum Hartle-Hawking states in JT gravity, I was previously able to relate the Bekenstein-Hawking area term for two-sided black holes in JT gravity to the log of a Plancherel measure for the semigroup $SL(2,R)^+$ \cite{Lin:2018xkj}. I will review this story in section \ref{s2} below (see especially eq. \eqref{main2}).

Here, I will extend this strategy to states set up by a Euclidean path integral with an operator inserted somewhere along the Euclidean boundary. Such states are known to be dual to wormholes in JT gravity with two horizons of different area \cite{Goel:2018ubv}. Moreover, the entanglement entropy between the boundary SYK models is known to equal the smaller of the two horizon areas to leading order in $1/G_N$. 
 I will show that this entanglement entropy can be written in a way that resembles a formula for entanglement entropy in a compact lattice gauge theory, where again the horizon areas in the JT description can be matched to Plancherel measures for the semigroup $SL(2,R)^+$, but where the lattice link that we appear to compute the entanglement  across depends on where we put the operator along the boundary. See section \ref{s31} for a summary of these results in equation form, which requires a little bit of notation to set up. I will then write down a family of density matrices in (an extension of) the compact gauge theory whose von Neumann entropies exhibit similar behavior as we tune an interpolating parameter. Finally, I will comment on the implications if we assume that a similar superselection sector structure as is responsible for the compact gauge theory result is present in gravity as well. The main implication is that each copy of the SYK model should then contain emergent $SL(2,R)$ intertwiners whose entanglement gives rise to space in the bulk.
 
 This paper is organized as follows. In section \ref{s2}, I will introduce the strategy by reviewing  \cite{Lin:2018xkj}. In section \ref{s3}, I will generalize the calculation of \cite{Lin:2018xkj} to states of JT gravity + matter that can be set up by a Euclidean path integral with an operator inserted along the Euclidean boundary, and show the results described above. 
 In section \ref{s4}, I will conclude with some comments on possible directions for future work. Finally, appendix \ref{sa} generalizes the calculation to multiple operator insertions, and appendix \ref{sb} contains some steps towards generalizing it to planar order.
	\begin{figure}
	\centering
	\begin{tabular}{ p{8.0cm} c p{7.7cm}}
	Gravity & & Compact gauge theory analog \\
	\begin{tcolorbox}
	1. Given a state set up by a Euclidean path integral in Lorentzian SYK/JT holography, write the entanglement entropy between the two boundary SYK models in a first order way. \end{tcolorbox} &  $\rightarrow$ & \begin{tcolorbox} 2a. In a compact lattice gauge theory, use the ``edge mode algorithm" (see section \ref{s22}) to write down a family of density matrices whose von Neumann entropies  have the same functional form. \end{tcolorbox} \\
	& & \hspace{30mm} $\downarrow$ \\
	\begin{tcolorbox} 3. Taking the gauge theory to be an analog of the {bulk}, 
	conjecture that there's a similar emergent
	structure in holography when we promote group representation labels $R$ to noncompact ones $k$, and see what this implies.\end{tcolorbox} & $\leftarrow$ & \begin{tcolorbox} (2b. Conservatively interpret the ``edge mode algorithm" as pointing to the presence of a certain superselection sector structure in the gauge theory.) \end{tcolorbox}
	\end{tabular}
	\caption{Summary of the strategy.}	
	\label{fig1} 
	\end{figure}

\section{Review of earlier results} \label{s2}

Since we will use the same strategy as in \cite{Lin:2018xkj}, let us start by reviewing that work.

\subsection{Setting and summary of earlier results}\label{s21}

The setting for \cite{Lin:2018xkj} was $1+1d$ Jackiw-Teitelboim (JT) gravity on an asymptotically AdS spacetime. This theory is described by the action 
\be\label{jtaction}
S = \frac{1}{16\pi G_N}\left[ \int_M d^2x \sqrt{-g}\, (R+2) \phi + 2\int_{\partial M}\phi\, K\right]\,,
\ee
where $\phi$ is a dynamical scalar field that we will call the dilaton, and $K$ is the extrinsic curvature of the boundary. The equations of motion that follow from \eqref{jtaction} are 
\begin{eqnarray}
(\nabla_\mu \nabla_\nu - g_{\mu\nu})\phi &=& 0\,, \label{jteom1} \\
R + 2 &=& 0\,, 	\label{jteom2}
\end{eqnarray}
the second of which sets the metric to be locally AdS. These equations should also be augmented by the boundary conditions 
\be\label{jtbc}
g|_{\partial M} = \frac{1}{\epsilon^2}\, \qquad \phi|_{\partial M}= \frac{\phi_b}{\epsilon}\,.
\ee
Below I will use the notation that $\alpha = (16\pi G_N)^{-1}$.

Eqs. \eqref{jteom1} - \eqref{jtbc} have a 1-parameter family of wormhole solutions \cite{Harlow:2018tqv}. Upon quantizing, these give rise to a continuous 1-dimensional Hilbert space. One natural basis for the Hilbert space is furnished by the Hartle-Hawking states, which are set up by a Euclidean path integral on a half-disk of length $\beta/2$. Another is provided by the index $k$ of principal continuous reps of $SL(2,R)$ \cite{Blommaert:2018oro}. See \cite{Harlow:2016vwg} for a more detailed discussion of JT gravity in Lorentzian signature and \cite{Maldacena:2016upp}, \cite{Engelsoy:2016xyb}, \cite{Jensen:2016pah} for earlier work on JT gravity.

An important fact about JT gravity is that it is the bulk side of an especially simple example of holography, being attainable in a double scaling (``Schwarzian") limit from two coupled SYK models in Lorentzian signature. (The SYK model is a $0+1d$ quantum-mechanical theory of interacting fermions with a random coupling; see \cite{Maldacena:2016hyu}, \cite{Kitaev:2017awl} for early work on the model.) The Euclidean path integral that sets up a Hartle-Hawking state in JT gravity sets up a parallel state in the boundary theory as well. Using standard methods, one can  show that the entanglement entropy between the two SYK models in this state can be written to leading order in $1/G_N$ as
\be\label{e25}
S \approx S_0 + \frac{\langle\phi_h\rangle_\beta}{4G_N}\,,
\ee
where $\langle\phi_h\rangle_\beta$ is the horizon value of the dilaton in the dual JT description, and  $S_0$ is a zero-point entropy which is independent of $\beta$. The former is the analog of the Bekenstein-Hawking area term in JT gravity.

In this setting, the main result of \cite{Lin:2018xkj} was that the entanglement entropy \eqref{e25}  between the two SYK models could alternatively be written as 
\be \label{main1}
S = S_0 + \int dk\, p_k(-\log p_k + \log k \sinh 2\pi k)\,,
\ee
where $k$ runs over the principal continuous reps of $SL(2,R)$, and the $p_k$'s are the norms of the Hartle-Hawking wavefunction coefficients in the representation basis of JT gravity. Moreover, at leading order in $1/G_N$, the second term in \eqref{main1} equals the Bekenstein-Hawking area term:
\be\label{main2}
\left.\log k \sinh 2\pi k\right|_{k = k_*(\beta)} = \frac{\langle\phi_h\rangle_\beta}{4G_N} = \left. \frac{A}{4G_N}\right|_{2d},
\ee
where $k_*(\beta)$ is the value of $k$ that extremizes $p_k$.  

The thing that we gain from writing the entanglement entropy in this way is that \eqref{main1} looks like a formula for entanglement entropy in a compact lattice gauge theory, where we can write down an exact density matrix that gives rise to the formula. That density matrix has a certain superselection sector structure, and if we assume that a similar superselection sector structure is responsible for \eqref{main1}, then we can draw some nice conclusions about SYK/JT holography. 
In the rest of this section, I will explain the details of this story in a slightly different way than the original presentation in  \cite{Lin:2018xkj}, by closely following the steps in Figure \ref{fig1}. 

\subsection{Step 1: Write the EE between the two SYK models in an unusual way}\label{s22}

According to Figure \ref{fig1}, the first thing that we should do is to exactly rewrite the entanglement entropy between the two SYK models in a useful way. To do this,  let us take
\be\label{e28}
p_k = Z_1^{-1}\, k \sinh 2\pi k\, e^{-\beta H(k)}\,,
\ee 
where $H = k^2/4\alpha\phi_b$ is the Hamiltonian of the SYK model in the Schwarzian limit (aka the Hamiltonian of JT gravity), and $Z_1$ is the partition function of Schwarzian QM (aka the partition function of JT gravity) on a circle of circumference $\beta$. For now, $p_k$ has no particular meaning beyond being the thing that will let us rewrite the entropy formula in a useful way. Where it comes from is that $Z_1p_k$ is the integrand of $Z_1$ in the representation basis of JT gravity with the integral stripped off. 

Given $p_k$, we can rewrite the quantities that appear in the replica trick formula in terms of it. According to the replica trick formula, the entanglement entropy of one of the two boundary QM's is given by
\be\label{replica}
S = \left. -\partial_n\left(\frac{Z_n}{Z_1^n}\right)\right|_{n=1} = -\left(\frac{\partial_n Z_n|_{n\rightarrow 1}}{Z_1}- \log Z_1\right)
\ee
where $Z_n$ is the $n$-replicated partition function for that tensor factor, here the Schwarzian partition function on a $n$-times-longer disk \cite{Stanford:2017thb}:
\be
Z_n = \int dk\, k \sinh 2\pi k\, e^{-n\beta H(k)}\,.
\ee
It's easy to see that we can rewrite this as
\be\label{e211}
Z_n = \int dk\, (k \sinh 2\pi k)^{1-n} (Z_1p_k)^n\,.
\ee
Eqs. \eqref{main1} and \eqref{main2} then follow from plugging \eqref{e211} into \eqref{replica}.

In short, the results of \cite{Lin:2018xkj} follow straightforwardly from exactly rewriting some well-known results for the Schwarzian partition function and related entropy formulas. So all of the nontrivial content will be  in how we interpret eqs. \eqref{main1} and \eqref{main2}.

\subsection{Step 2: Find a family of density matrices whose von Neumann entropies are structurally identical to \eqref{main1}}

To get started on this, we next find a family of density matrices in a non-gravitational quantum system whose von Neumann entropies are structurally identical to \eqref{main1}. 

The quantum system in question will be a compact gauge theory whose Hilbert space I will extend via a procedure that I'll call the ``edge mode algorithm". Before getting started, let me stress for expert readers that this algorithm {\it cannot} be applied directly to gravity for reasons that I'll come back to in section \ref{s4}, and I will not be able to write down an exact density matrix whose von Neumann entropy equals \eqref{main1}. However, the compact gauge theory result can more weakly be interpreted as pointing to the presence of a particular superselection sector structure in the problem, 
and {that} aspect of it alone is what I will try to generalize to gravity in the later sections.

\subsubsection{Definition of the edge mode algorithm}\label{s231}

More specifically, consider pure 2d Yang-Mills theory on $S^1$ with a compact gauge group $G$. An orthonormal basis for its Hilbert space is furnished by representations $R$ of the gauge group $G$ \cite{Cordes:1994fc}. Given some state $\psi$ in this theory, suppose that we want to assign an entanglement entropy to an interval $I$ of the spatial $S^1$. One problem that we immediately run into is that the Hilbert space doesn't factorize, due to the requirement from Gauss's law that the electric flux through the boundary of $I$ has to match the flux through the boundary of its complement $\bar I$. This precludes us from tracing out $\bar I$ to assign a reduced density matrix to $I$.

To get around this, one rather naive thing that we can do is to embed the Hilbert space of the gauge theory into a larger one,
\be\label{e212}
\HH_{phys} \rightarrow  \HH_{ext.} = \HH_{I} \otimes \HH_{\bar{I}}\,,
\ee
where we add in degrees of freedom that transform under the gauge group at the boundaries of $I$ \cite{Donnelly:2014gva}. 
In practice, we can implement this by adding pairs of lattice sites to the ends of $I$ that we populate with non-dynamical surface charges in all possible representations of $G$. When we embed each state $|R\rangle$ from the original Hilbert space $\HH_{phys}$ into the extended one, Gauss's law requires that we pair up surface charges along the cut, \footnote{I.e. $|R\rangle \rightarrow$ $|R,i,j\rangle \otimes |R,j,i\rangle$, up to normalization.} so the embedding will actually pick out a subspace of $\HH_{ext.}$ consisting of superselection sectors labeled by $R$: 
\be\label{e213}
\HH_{phys} \rightarrow \bigoplus_R \HH_I^R \otimes \HH_{\bar I}^R\subset \HH_{ext.}\,.
\ee

Given the state $\psi(R)|R\rangle$, suppose that we now want to compute the entanglement entropy of the interval $I$ {\it w.r.t. this particular extension of the Hilbert space}. After embedding $\psi$ into $\HH_{ext.}$ as 
\be
\psi(R)|R\rangle \, \rightarrow \,  \psi(R)\,(\dim R)^{-1} |R,i,j\rangle\otimes |R,j,i\rangle\,,
\ee
we find that the reduced density matrix for one of the tensor factors is 
\be\label{rhor}
\rho_I = \bigoplus_R  \frac{p_R\bone_R}{(\dim R)^2}\,,
\ee
where $p_R = |\psi(R)|^2$, and ${\bf 1}_R$ is the identity matrix of size $(\dim R)^2$. The von Neumann entropy of $\rho_I$ is 
\be\label{lgtee}
S = \sum_R p_R (- \log p_R + 2\log \dim R)\,,
\ee
resembling \eqref{main1}. \footnote{Up to a factor of 2 that comes from there being two endpoints to the entangling interval on $S^1$.} In short, if we apply the edge mode algorithm to the problem of assigning entanglement entropy to subregions in a compact $1+1d$ Yang-Mills gauge theory, we find a family of exact density matrices in an extension of the compact gauge theory whose von Neumann entropies take the same form as our unusual way of writing the gravitational entropy formula, \eqref{main1}.

This edge mode algorithm can be generalized to give a definition of entanglement entropy for any subregion of a lattice gauge theory in any number of dimensions \cite{Buividovich:2009zz}, \cite{Donnelly:2011hn}, \cite{Ghosh:2015iwa}. In $1+1d$, its application to an arbitrarily lattice yields 
\be\label{lgtee2}
S = \sum_{R_\partial} p(R_\partial) (-\log p({R_\partial}) +  \sum_{e \in \partial} \log \dim R_e)\,, \ee
where $p(R_\partial)$ is the probability distribution of the representations on the boundary links only, and $e \in \partial$ runs over the boundary links. Eq. \eqref{lgtee} is a special case of this formula when there is just one lattice link. We will use eq. \eqref{lgtee2} in section \ref{s3} below.

\subsubsection{The meaning of the edge mode algorithm} 

While the steps described above are {technically} well-defined for any compact lattice gauge theory, the edge mode algorithm might seem poorly motivated from a physical point of view. After all, we're taught that gauge symmetries are redundancies in the description of a theory, so putting in gauge-variant degrees of freedom by hand might seem to be in bad taste. Let's therefore see if we can understand eq. \eqref{lgtee} from a purely gauge-invariant point of view.

The first term in \eqref{lgtee} indeed has an information-theoretic interpretation within the gauge theory as the algebraic entanglement entropy of gauge-invariant operators contained entirely within the inteval $I$. (The entanglement entropy of an algebra $\mathcal{A}$ in a state $\psi$ is defined to be the von Neumann entropy of the unique operator $\rho \in \mathcal{A}$ such that $\Tr \rho\, \OO = \langle \psi|\OO|\psi\rangle$, for all $\OO \in \mathcal{A}$ \cite{Casini:2013rba}.) In this example, the gauge-invariant operators contained within $I$ consist entirely of the Casimirs built from the electric field operator, so the algebraic entanglement entropy coincides with a classical Shannon entropy for the superselection sectors labeled by $R$.

The $``\log \dim R$" term, on the other hand, doesn't have an operational interpretation within the gauge theory. Instead, it's just a label that we assign to each state $|R\rangle$. 
(This shouldn't be surprising since it is clearly counting the entanglement between the fictitious surface charges when we extend the Hilbert space.)
One can argue that such a term must always be present if the gauge theory emerges from a more UV theory on the lattice w.r.t. which we compute the entanglement entropy \cite{Lin:2017uzr}, and there seems to be a more general sense in which such terms should always
	appear if we try to define entanglement entropy in a theory with constraints, in order to quantify the size of the phase space orbit that we would get from lifting the constraints between the subsystems (see e.g. \cite{Riello:2021lfl}). However, no general derivation of how this edge term follows from a given constraint equation has been worked out, as far as I know. 


In this paper, I will take the more conservative view that the appearance of a $``\log$ (group-theoretic measure labeled by $k$)" term in an entanglement entropy formula just tells us {\it qualitatively} that the underlying Hilbert space contains superselection sectors labeled by $k$. I will not attempt to give a microscopic explanation for why the $``\log \dim R"$ term has the exact value that it does. In other words, throughout the paper I will try to see how much mileage we can get by replacing things labeled by $R$ on the compact gauge theory side of the analogy to things labeled by $k$ on the gravity side, but I will not attempt to do anything with the magnetic indices.
	
\subsection{Step 3: Interpret the holographic formula \eqref{main1} by analogy} \label{s24}

Let us now see what happens if we assume that the gravitational entropy formula \eqref{main1}  comes from a similar superselection sector structure \eqref{e213} as eq. \eqref{lgtee} in the compact gauge theory. From this assumption, it follows that

\begin{itemize}
\item The theory of the two coupled SYK models should have states labeled by $SL(2,R)$ representations $k$ in the Schwarzian limit;  
\item 
From comparison to \eqref{e213}, 
{\it each} of the SYK models should itself contain a Hilbert space sector that in some sense is labeled by $k$ in the Schwarzian limit, \footnote{Or equivalently, 
composite operators labeled by $k$ that can be used to reach states in that Hilbert space sector starting from the ground state, since SYK itself has no superselection sectors; they appear only when we take the gravity limit.} 
although I don't understand the size of this Hilbert space sector or structure of states within it because I don't understand the canonical origin of the $k \sinh 2\pi k$ density of states (henceforth ``the sinh"); 
and
\item In terms of this label $k$, the Bekenstein-Hawking area term in the JT description follows from eq. \eqref{main2}.
\end{itemize}

In this vacuum example, we basically already knew that the first two things were true from the representation basis of JT gravity \cite{Blommaert:2018oro} and the $G\Sigma$ variables,
 and we can now check that the $p_k$'s defined in an ad hoc way in \eqref{e28} are in fact the square of the overlap of the Hartle-Hawking wavefunction with states in the representation basis of JT gravity. 

Note that these points fit nicely with the idea of 
bit threads \cite{Freedman:2016zud} and its refinement in \cite{Casini:2019kex}. In more conventional examples of holography, \cite{Freedman:2016zud} observed that the Ryu-Takayanagi area term assigned to a given boundary region in a holographic CFT can be obtained by maximizing a surface integral of divergenceless vector fields that emanate from that region with a fixed maximal density per unit area. The RT area term then arises from the tightest possible packing of flux lines for the vector field, i.e. the eponymous bit threads, in the bulk. Though this observation is suggestive, pointing to a pre-geometric picture of quantum gravity where space itself consists of objects that create and prop open the space, it leaves open the  question of {\it what} the bit threads are supposed to be. 
In \cite{Casini:2019kex}, the authors suggested a partial answer (from applying algebraic QFT arguments to the entanglement wedge structure of the bulk effective field theory): that the bit threads might be the bulk duals of high-dimension charge-neutral composite operators in the boundary theory, made up of charged operators living in different spatial regions on the boundary. \footnote{In their paper, they call these charge-neutral operators that span boundary regions ``intertwiners". These should not be confused with the charged intertwiners that I will discuss in section \ref{s34}, which will live within one of the two SYK's. The analog of their intertwiners are the bilocals formed from what I will call intertwiners in section \ref{s34}.
		} 		
In our picture, each state in the representation basis of JT gravity is like a ``single bit thread" that explicitly realizes their idea. We will further refine this picture in the examples below.

\section{Entanglement entropy in JT gravity with matter} \label{s3}  

I will now generalize the previous calculation to the case where we include an operator  along the boundary of the Euclidean path integral that sets up the Hartle-Hawking state. This will let us study an entanglement phase transition that appears as we tune the location of the operator. 

\subsection{Setting and summary of results} \label{s31}

\begin{figure}
	\centering
	\includegraphics[height=1.5in]{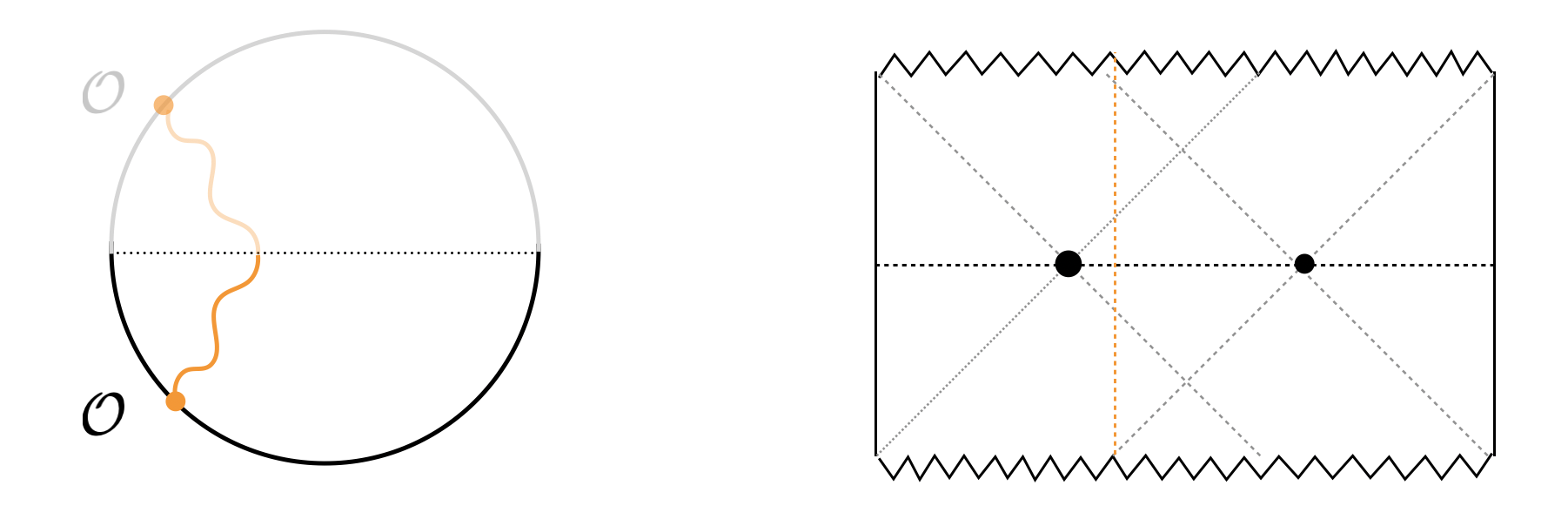}
	\caption{Left: The states that we will study in this section are obtained by cutting the two-point function in \eqref{e31} at the moment of time-reflection symmetry. Right: They correspond in the bulk to two-sided wormholes with two horizons of different area.}
	\label{fig2} 
	\end{figure}

More specifically, the states that we now want to consider are the ones obtained by cutting a Schwarzian two-point function \cite{Mertens:2017mtv}
\be \label{e31}
\langle\OO(\tau) \OO(0)\rangle_\beta = \int \prod_{i=1,2} dk_i\, k_i \sinh 2\pi k_i\, e^{-\frac{k_1^2}{2C}\tau}e^{-\frac{k_2^2}{2C}(\beta - \tau)}\frac{\Gamma(\ell \pm ik_1 \pm ik_2)}{\Gamma(2\ell)}
\ee
at the moment of time-reflection symmetry, where $C = 2\alpha\phi_b$ in our earlier notation, and the ``$\pm$" inside the gamma function means that we should take the product over all possible combinations of the $\pm$'s. These states were studied by \cite{Goel:2018ubv}, who showed that at leading order in $1/G_N$ and for operator dimensions $\Delta(\OO) = \ell$ of order $1/G_N$, they describe wormholes in the bulk with two horizons of different area (see Figure \ref{fig2}). Moreover, they showed that at leading order in $1/G_N$ and away from $\tau = \beta/2$, 
	the entanglement entropy between the two SYK models turns out to be the smaller of the two horizon areas: 
\be\label{e32}
S = S_0 + \frac{1}{4G_N} \min(\langle\phi_L\rangle_{\tau, \beta}, \langle\phi_R\rangle_{\tau, \beta})\,.
\ee

In this section, we will find that the entanglement entropy between the two SYK models can instead be written as 
\be\label{main21} 
\boxed{S = S_0 + 
\min_{i = 1,2}\left(\int dk_i\, p_{k_i}(- \log p_{k_i} + \log k_i \sinh 2\pi k_i) \right)}
\ee
to leading order in $1/G_N$, where 
\begin{eqnarray}
p_{k_1} &=& \int dk_2\, p_{k_1k_2}\,, \label{e34} \\
p_{k_2} &=& \int dk_1\, p_{k_1k_2}\,, \label{e35} \\
\label{e36}
p_{k_1k_2} &=& Z_1^{-1}\, k_1 \sinh 2\pi k_1\, k_2 \sinh 2\pi k_2\, e^{-\frac{k_1^2}{2C}\tau}e^{-\frac{k_2^2}{2C}(\beta - \tau)}\frac{\Gamma(\ell \pm ik_1 \pm ik_2)}{\Gamma(2\ell)}\,,
\end{eqnarray}
and throughout the section we will take $Z_1$ to be $\langle\OO(\tau)\OO(0)\rangle_\beta$, \eqref{e31}.
Moreover, as before, the Bekenstein-Hawking area terms in \eqref{e32} can be obtained by evaluating the $``\log \dim R"$ terms in \eqref{main21} at the peak of the distributions defined by $p_{k_1}$ and $p_{k_2}$: 
\be\label{main22}
\boxed{\left. \log k_1 \sinh 2\pi k_1\right|_{k_1 = k_1^*(\tau, \beta)} = \frac{\langle\phi_L\rangle_{\tau, \beta}}{4G_N}}\,,
\ee
\be
\boxed{\left. \log k_2 \sinh 2\pi k_2\right|_{k_2 = k_2^*(\tau, \beta)} = \label{main222}\frac{\langle\phi_R\rangle_{\tau, \beta}}{4G_N}}\,,
\ee
where $k_1^*(\tau, \beta)$, $k_2^*(\tau, \beta)$ are the values of the $k_i$'s that extremize \eqref{e34} and \eqref{e35}. 

The interesting thing about this way of writing the entropy formula is that the right-hand side of \eqref{main21} is exactly like the formula that one finds when computing entanglement entropy in a lattice gauge theory on a lattice with two links using the ``edge mode algorithm",  \eqref{lgtee2}, {but where the link that we compute the entanglement entropy across, naively the choice of tensor factor that we trace out, depends on the outcome of the minimization step}. Yet at the same time, \eqref{main21} came from an honest calculation in the boundary theory where we traced out one of two fixed tensor factors in the boundary Hilbert space. This motivates the question of whether we can find a parametric family of states in a compact gauge theory example where we genuinely get a phase transition in the von Neumann entropy that looks like \eqref{main21} as we tune an interpolating parameter, while keeping the location of the entanglement cut fixed.

In section \ref{s33}, I will write down a toy model with this property. The key ingredient will be that entanglement entropy in gauge theories is topological w.r.t. flux lines and doesn't care where along a flux line we make the cut. If an analogous mechanism is responsible for \eqref{main21}, it could explain some things around how SYK/JT holography works. I will describe some consequences of taking the analogy seriously in sections \ref{s34} and \ref{s4}. 

\subsection{Step 1: Write the EE between the two SYK models in an unusual way}\label{s32}

Let us start by deriving eqs. \eqref{main21} - \eqref{main222}. The steps here will again be totally mechanical and the reader who is interested in the meaning of the results can skip ahead to section \ref{s33}. 

\begin{figure}
	\centering
	\includegraphics[height=1.4in]{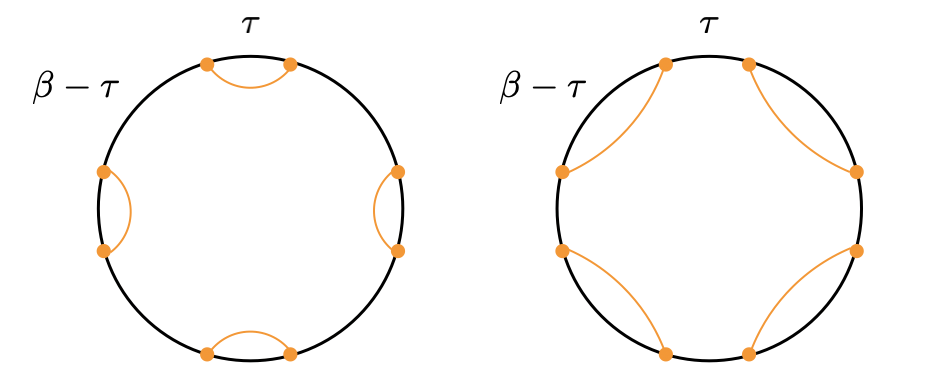}
	\caption{Illustration of the leading channel that contributes to $Z_n$ for $n=4$ when $\tau < \beta/2$ (left) and $\tau > \beta/2$ (right). This figure is not drawn to scale.}
	\label{fig3} 
	\end{figure}

As in \cite{Lin:2018xkj}, our strategy will be to rewrite the quantities that appear in the replica trick formula \eqref{replica} in terms of $p_{k_1k_2}$, \eqref{e36}. In this case, the $n$-replicated partition function $Z_n$ for the entanglement entropy between the two SYK models in the Schwarzian limit will be the Schwarzian partition function on a thermal circle of circumference $n\beta$ with $2n$ operator insertions. If we wanted to compute this partition function exactly, we would have to sum over all ways of contracting the operators, which we don't know how to do. To leading order in $1/G_N$ though, we can keep just the dominant channel, which is the one where we pairwise contract the operators that are closest to each other along the thermal circle (for $\tau \neq \beta/2$) \cite{Goel:2018ubv}; see Figure \ref{fig3}. \footnote{When $\tau = \beta/2$ there is a degeneracy in the dominant channel, so to use the replica trick, we would have to find some subset of contributions to $Z_n$ that includes both channels and has a good analytic continuation in $n$. See Appendix \ref{sb} for some steps in this direction.}

Without loss of generality, let us specialize to $\tau < \beta/2$. In this case, the dominant channel that contributes to $Z_n$ evaluates to 
\be
	Z_n \approx \int\dots \int \prod_{i=1\dots n} dk_i\, k_i \sinh 2\pi k_i\, e^{-\frac{k_i^2}{2C}\tau}\int dk_{b}\, k_{b} \sinh 2\pi k_{b}\, e^{-n\frac{k_{b}^2}{2C}(\beta - \tau)} \frac{\Gamma(\ell \pm ik_i \pm ik_{b})}{\Gamma(2\ell)}\,,
\ee
where $k_b$ is the integration variable that we assign to the region that connects different boundary segments in the bulk. (See \cite{Blommaert:2018oro} for the Feynman rules that go into computing these diagrams.) In terms of $p_{k_1k_2}$ that we defined in \eqref{e36}, we can rewrite this as 
\begin{eqnarray}
	Z_n &\approx& \int d{k_{b}}\, (k_{b}\sinh 2\pi k_{b})^{1-n}\left[\int dk\,  k \sinh 2\pi k \, e^{-\frac{k^2}{2C}\tau}\,k_{b}\sinh 2\pi k_{b} \, e^{-\frac{k^2_{b}}{2C}(\beta - \tau)} \, \frac{\Gamma(\ell \pm ik \pm i k_{b})}{\Gamma(2\ell)}\right]^n \\
	&\approx& \int dk_2 \, (k_2 \sinh 2\pi k_2)^{1-n}\left[\int dk_1\, Z_1 p_{k_1k_2}\right]^n\,.\label{e311}
\end{eqnarray}
Plugging \eqref{e311} into the replica trick formula \eqref{replica} then yields \eqref{main21} with $i=2$. The derivation for $\tau > \beta/2$ is similar.

Next, we'd like to show that the Bekenstein-Hawking area terms in \eqref{e32} follow from evaluating the ``$\log \dim R"$ terms in \eqref{main21} at the peaks of the distributions defined by $p_{k_1}$ and $p_{k_2}$. To do this, we will use the fact from \cite{Goel:2018ubv} that the area terms 
are related to the solutions $k_{1, s.p.}$, $k_{2, s.p.}$ of the saddle-point equations for the integrand of $\eqref{e31}$, 
\begin{eqnarray}\label{sp1}
\frac{k_1\tau}{C} + 2 \arctan \frac{k_1 + k_2}{\ell} + 2\arctan\frac{k_1-k_2}{\ell}	 &=& 2\pi, \\
\frac{k_2\cdot (\beta - \tau)}{C} + 2\arctan\frac{k_1 + k_2}{\ell} - 2\arctan\frac{k_1 - k_2}{\ell}\,, &=& 2\pi\,,\label{sp2}
\end{eqnarray}
as
\begin{eqnarray}\label{e314}
\frac{\langle\phi_L\rangle}{4G_N} = 2\pi k_{2, s.p.}, \qquad \frac{\langle\phi_R\rangle}{4G_N} = 2\pi k_{1, s.p.}\,.
\end{eqnarray}
(Note that both sides of \eqref{e314} implicitly depend on $\tau$ and $\beta$.) So to prove \eqref{main22} and \eqref{main222}, we just have to show that 
\be\label{e315}
\lim_{k \rightarrow \infty} \left. \log k_1 \sinh 2\pi k_1\right|_{k_1 = k_1^*(\tau, \beta)} = 2\pi k_1^*(\tau, \beta) = 2\pi k_{1, s.p.}(\tau, \beta)
\ee
for $k_1^*(\tau, \beta)$ the value of $k_1$ that extremizes \eqref{e34}, and similarly,
\be\label{e316}
  2\pi k_2^*(\tau, \beta) = 2\pi k_{2, s.p.}(\tau, \beta)\,.
\ee
But this almost follows from the fact that $p_{k_1k_2}$ was defined as the very integrand of \eqref{e31} from which the saddle-point equations \eqref{sp1} and \eqref{sp2} were drawn. The only nontrivial step left to account for is that \eqref{e34} and \eqref{e35} each involve marginalizing over one of the two $k_i$'s, but this won't change the extremum of the other if the joint distribution is sufficiently sharply peaked. 

Note that while \eqref{e315} and \eqref{e316} may look rather trivial, the $2\pi$'s on either side of the equations came from different places. In the thermodynamic approach, they came from a cancellation of coefficients between the free energy and vev of the Hamiltonian, while in our approach they came directly from the large $k$ limit of the measure $``k \sinh 2\pi k$".

To summarize, in this section we generalized the calculation of section \ref{s22} to states with an operator insertion in the path integral, rewriting the results of \cite{Goel:2018ubv} in a first order way. We would now like to interpret \eqref{main21} by finding an analogous  formula in a compact gauge theory situation where we know the microscopic meaning of all of the terms in the formula. In other words, we would like to find a family of exact density matrices whose von Neumann entropies in some limit (the analog of the $G_N \rightarrow 0$ limit) interpolate between expressions like \eqref{main21} with $i=1$ and $i=2$ as we tune an external parameter.

\subsection{Step 2: Find a family of density matrices whose von Neumann entropies are structurally identical to \eqref{main21}}\label{s33}


As mentioned already, a starting point for our search is that \eqref{main21} for fixed $i$ resembles the ``edge mode" definition of the entanglement entropy across a link in a lattice gauge theory, \eqref{lgtee2}. However, $i$ is not fixed in \eqref{main21}, but depends on the outcome of the minimization step. So the challenge is whether we can get a formula like it while tracing out a fixed subspace of a Hilbert space. 

\begin{figure}
	\centering
	\includegraphics[height=0.8in]{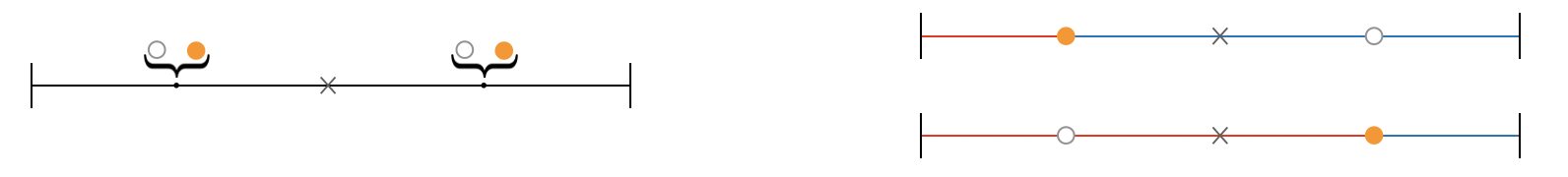}
	\caption{Left: In this section, we consider the entanglement entropy across the location marked ``x" in a lattice gauge theory with three links and two nodes, each of which contains either no charge or a static charge in a representation {\bf R} (depicted as an orange node in the picture). Right: Moreover, we specialize to the subspace of states where the middle link is in the same representation as either the link to its left or the link to its right.}
	\label{fig4} 
\end{figure}

To do this, the trick is to notice that entanglement entropy in a lattice gauge theory is topological w.r.t. a given flux line, not caring where along the flux line we make the cut. This suggests the following toy situation. Suppose that we have a lattice gauge theory on a lattice with three links and two nodes, each of which contains either no charge (aka a charge in the identity representation) or a static charge in some representation ${\bf R}$ that mimics the quantum numbers of our operator $\OO$. \footnote{See section 2.6 of \cite{knowles} for a review of lattice gauge theory with charges at the lattice sites.} 
Moreover, let us specialize to the subspace of states in the lattice gauge theory where the middle link is in either the same rep $R$ as the link to its left or the same rep $R' \neq R$ as the link to its right. 
We can parametrize such states as
\be\begin{split}\label{e317}
|\psi\rangle &= \sum_{R,i,j,R',i',j'} \psi_1(R,i,j,R',i',j')\,\,|R,i,j \rangle \otimes |R',i',k'\rangle \otimes |R',k',j'\rangle \\ &+ \sum_{R,i,j,R',i',j'} \psi_2(R,i,j, R',i',j')\,\,|R,i,k\rangle \otimes |R,k,j\rangle \otimes |R',i',j'\rangle\,,
\end{split}\ee	
where the $\psi_{a}(R,i,j,R',i',j')$'s are arbitrary wavefunction coefficients, $i \in 1, \dots, \dim R$ and $i' \in 1, \dots, \dim R'$ are the magnetic quantum numbers at the ends of the links, and all repeated indices that aren't explicitly included in the sum should also be  implicitly summed over.
See Figure \ref{fig4} for a drawing of the situation.  

Now suppose that we want to compute the entanglement entropy across the middle of the three links (the location marked ``x" in Figure \ref{fig4}). As before, we face the problem that the gauge theory doesn't factorize across the cut, but we can brute-force our way around it by introducing a new lattice node at ``x"  that we populate with static charges in all possible representations of the gauge group. 
In the extended Hilbert space containing the new node, which has the structure  
\be\label{e318}
\HH_{phys} \subset \bigoplus_{R, R'} \HH_I^{R, R'} \otimes \HH_{\bar I}^{R'} \oplus \HH_I^R \otimes \HH_{\bar I}^{R, R'} 
\subset \HH_{link}^{\otimes 4}\,, \ee 
 $|\psi\rangle$ gets promoted to
\be\begin{split}\label{e3182}
|\psi\rangle &= \sum_{R,i,j,R',i',j'}  \psi_{1} (R,i,j, R',i',j') \,\,|R,i,j\rangle \otimes |R',i',\ell'\rangle \otimes |R'\,\ell', k'\rangle  \otimes |R',k',j'\rangle\\ &+ \sum_{R,i,j,R',i',j'}  \psi_2(R,i,j, R',i',j')\,\,|R,i,k\rangle \otimes |R,k,\ell\rangle \otimes |R,\ell,j\rangle \otimes |R',i',j'\rangle\,,
\end{split}\ee
up to normalization. Let us assume that \eqref{e3182} has been normalized to 1.

We can now trace out the first two tensor factors in \eqref{e3182}. When we do this, two things happen. First, 
 the reduced density matrix splits up as 
\be \rho = \bigoplus_{a=1,2} p_a\rho_a\,,
\ee
where 
\begin{eqnarray}
p_1 &=& \sum_{R,i,j,R',i',j'}|\psi_1(R,i,j,R',i',j')|^2\,, \\
p_2 &=& \sum_{R,i,j,R',i',j'}|\psi_2(R,i,j,R',i',j')|^2\,.
\end{eqnarray}
Within each of the two sectors, the situation is then the same as in \eqref{lgtee2} for a fixed selection of boundary links. We  get a Shannon entropy for the probability distribution over the reps on the link that the cut goes through, and a $``\log \dim R"$ term from pairing fictitious edge modes across the cut. Altogether, this leads to the result
\be\label{e320} 
S = - \sum_{a = 1,2} p_a \log p_a + p_1\,p_{Ri}\sum_{R,i} (- \log p_{R,i} + \log \dim R) +  p_2\, p_{R'j'} \sum_{R',j'}(-\log p_{R',j'} + \log \dim R')
\ee
for the entanglement entropy across the middle link, 
where
\begin{eqnarray}
p_{Ri} &=& \sum_{R',i',j,j'} |\psi_1(R,i,j,R',i',j')|^2\,, \\
p_{R'j'} &=& \sum_{R,i,i',j}\,\,\,\, |\psi_2(R,i,j,R',i',j')|^2	\,.
\end{eqnarray} 
If we further specialize to a family of states s.t. $p_1$ and $p_2$ are approximately oppositely-oriented step functions, then \eqref{e320} has almost the same form as the gravity result \eqref{main21} that we set out to explain. 
The only difference between \eqref{e320} for $p_1$ and $p_2$ of this form and \eqref{main21} is that \eqref{e320} has some extra magnetic indices, but those are a boring artifact of working on a lattice with a boundary. In the JT gravity case, the magnetic indices at the AdS boundary are removed by Hamiltonian reduction \cite{Blommaert:2018oro}. 


To summarize, the key ingredient that we needed to get the  formula \eqref{e320} in lattice gauge theory was to have two sectors where the link that the entanglement cut goes through is in either the same representation as the link to its left or the link to its right. This allowed us to mimic a situation where the entanglement cut goes through different flux lines.

\subsection{Step 3: Interpret the holographic results by analogy} \label{s34}


Let us now see what happens if we assume that \eqref{main21} comes from a similar superselection sector structure, \eqref{e318},  as \eqref{e320}. As before, I will only assume that we can replace compact group representation labels $R$ with $SL(2,R)$ representation labels $k$, and will not attempt to generalize the magnetic indices or explain what the analog of the  $``\log \dim R"$ term in gravity is counting. This section will nonetheless be a fair bit less precise than the rest, but I'll discuss how we might try to put it on firmer footing in section \ref{s4}.

With this assumption, our analogy suggests that

\begin{itemize}
\item The theory of two coupled SYK models should have states 
labeled by pairs of $SL(2,R)$ representations $k_1$ and $k_2$ in the Schwarzian limit.

\item In order for the mechanism that gives rise to  \eqref{e320} to explain \eqref{main21}, {\it each} copy of the SYK model should {itself} contain a subspace of 
states that in some sense are naturally labeled by two $k_i$'s 
in the Schwarzian limit, although I don't know exactly what the subspaces should look like because I don't understand the microscopic origin of the sinh. 
Equivalently, the SYK model should contain collections of composite operators labeled by two $k_i$'s that can be used to reach those Hilbert space sectors starting from the ground state. I'll schematically draw both the Hilbert space sectors and operators suggested by the analogy as \includegraphics[height=0.2in]{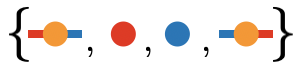}, with different colors standing for different representations, where hopefully the meaning will be clear from the context. 
\item These operators can again be combined to form gauge-invariant bilocal operators in the Schwarzian limit, in a way that again I don't know the details of because I don't understand the origin of the sinh. Interestingly though, perhaps we can construct a given bilocal operator in multiple ways. E.g. \includegraphics[height=0.22in]{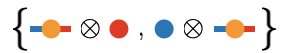} seem to describe the same thing in the IR. 

\item In this schematic notation, a state in the boundary theory set up by a Euclidean path integral with an operator inserted somewhere along the Euclidean boundary will have the form
\be\label{e325}
\includegraphics[height=0.3in]{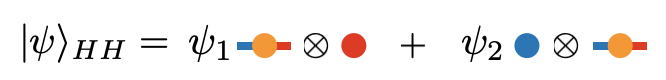}\,,
\ee
where each term on the right-hand side represents an entire superselection sector, and the $\psi_a$'s are such that either the sector labeled by $a=1$ or the one labeled by $a=2$ will have most of the probability mass for a given choice of the operator's location along the Euclidean boundary (away from $\tau = \beta/2$).


\item In terms of $k_1$ or $k_2$, the Bekenstein-Hawking area term in the JT description follows from eqs. \eqref{main22} and \eqref{main222}.
\end{itemize}
Again, these things follow from {\it assuming} that \eqref{main21} comes from similar underlying physics as \eqref{e320}, so at the moment they are conjectures (except for point \#5, which we explicitly showed above).

Some of these things are basically already known while others are new. The status of these points is as follows:

The first point is consistent with the boundary particle formalism \cite{Maldacena:2016upp} when the matter sector consists of just one particle. Namely, in pure JT gravity there are no propagating modes since the dilaton equation of motion \eqref{jteom2} sets the bulk to be locally AdS$_2$; the only gravitational degrees of freedom are the locations of the physical boundaries, which are not coupled directly to each other, but are coupled indirectly through a $SL(2,R)$ symmetry; and this remains true when we add matter, but the matter can also carry a $SL(2,R)$ charge. Hence, we can write the Hilbert space of JT gravity + matter as (see the discussion around eq. (2.15) of \cite{Lin:2019qwu})
	\be \label{hguess1}
	\HH = (\HH_L \otimes \HH_m \otimes \HH_R)/SL(2)\,, \qquad Q^a_L + Q^a_m + Q^a_R = 0\,,
	\ee
where the $Q^a$'s are the $SL(2,R)$ charges of the individual systems. If we specialize to states where the ``matter sector" consists of one heavy particle, this gives the basis that we want.
\footnote{Turning things around, consistency with the boundary particle formalism also allows us to privilege the toy model in section \ref{s33} over more complicated ansatze that in principle could lead to \eqref{e320}. For example, suppose that rather than starting from the Hilbert space $\HH_{link}^{\otimes 3}$ of a lattice gauge theory in section \ref{s33}, we started from a larger Hilbert space $(\HH_{link}\oplus |\Delta\rangle)^{\otimes 3}$ 
and considered states in the subspace spanned by $\{|\Delta\rangle \otimes |R',i',k'\rangle \otimes |R',k',j'\rangle, |R,i,k\rangle\otimes |R,k,j\rangle \otimes |\Delta\}$ in place of \eqref{e317}. Then we could still get to something like like \eqref{e320} from following similar steps, but we'd have to explain what the analogs of the $|\Delta\rangle$'s are supposed to be in JT gravity, if not $SL(2,R)$ representation labels.}

The second point I think is new 
and is the main conceptual point of this paper. However, it is rather vague since I have not explicitly defined what these objects are supposed to be that intertwine $SL(2,R)$ representations within each copy of the SYK model. I will come back to this in section \ref{s4}.

The third and fourth points are naive consequences of the second. 


Finally, the fifth point is also new (and is along with the analogy between \eqref{main21} and \eqref{e320} our main justification for the speculative 2nd point), but is a natural follow-up to the result of \cite{Lin:2018xkj}.

Comments:


\begin{itemize}
\item[(*)] 	The combination of the first and fifth points suggests a picture of the bulk where space itself consists of a dynamical network whose link quantum numbers give rise to areas. This is reminiscent of the situation in loop quantum gravity (see \cite{Rovelli:2014ssa} for a review). However, the main point of this section is not this (IR) story, but the way that its holographic encoding may have a redundant form like in \eqref{e325}.

\item[(*)] It's also interesting to think about what these points would imply for the black hole information paradox if we make the additional assumption that the entanglement phase transition in \eqref{main21} and in the Island formula \eqref{islandrule} are of a similar kind. Since the states that we studied here are formally analogous to the ones in the ``JT gravity + end-of-the-world brane" toy model of \cite{Penington:2019kki} if we identify their radiation system $\mathsf{R}$ with one of our two copies of the SYK model, this assumption is the usual ``ER = EPR" one that two-sided and one-sided black holes work the same way \cite{Maldacena:2013xja}. 
If we make this assumption, we seem to find that an important ingredient for the appearance of the Island formula is that there are multiple ways to encode the same state in the bulk effective field theory into a particular partition of the boundary theory into tensor factors, one for each additional non-minimal extremal surface, 
and that Hawking's original paradox
comes from neglecting the extra encodings and sticking to just
one of them over the course of the black hole lifetime. But it is difficult to say more about
this without having a more explicit construction to work with. \end{itemize}

\section{Discussion} \label{s4}

To summarize, in this paper I first showed that in a state of JT gravity set up by a path integral with an operator inserted somewhere along the Euclidean boundary, the entanglement entropy between the two boundary SYK models 
  can be written to leading order in $1/G_N$ in a way that resembles a formula for entanglement entropy in a compact lattice gauge theory, but where the link that we compute the entanglement across depends on where we put the operator along the boundary, \eqref{main21}. Moreover, the Bekenstein-Hawking area term in this situation can be written as the log of a Plancherel measure for the semigroup $SL(2,R)^+$, \eqref{main22}, \eqref{main222}. These were our main technical results,  as summarized in section \ref{s31}.

I then tried to interpret the resul by finding
an explicit family of density matrices in (an extension of) a compact lattice gauge theory whose von Neumann entropies \eqref{e320} had a similar form as \eqref{main21}. The key mechanism 
needed for this  to work was that the entanglement cut had to effectively go through different flux lines of a lattice with a fixed underlying topology. If the mechanism leading to \eqref{main21} is the same, it suggests that {\it each} copy of the SYK model should contain emergent Hilbert space sectors labeled by multiple $SL(2,R)$ representations, as discussed in section \ref{s34}. 

There are a number of follow-up topics that we would like to understand better. 

The most important is whether the speculations in section \ref{s34} are correct. There, we conjectured that the SYK model contains Hilbert space sectors labeled by multiple $SL(2,R)$ representations in the double scaling limit. Can we explicitly find this structure, write down an effective action for the $SL(2,R)$ intertwiners and understand how to go from it to the JT gravity action in the bulk? This seems hard since it would basically amount to solving SYK/JT holography.


An important intermediate step is to understand how to get the ``$k \sinh 2\pi k$" density of states from the SYK model. Throughout this paper, I have tried to emphasize that when comparing JT gravity to a compact gauge theory, I have only conjectured that the superselection sector structure generalizes when we replace things labeled by representations $R$ of the compact gauge group with things labeled by representations $k$ of $SL(2,R)$. I don't know how to generalize the magnetic indices or where the $``k \sinh 2\pi k"$ comes from.
If it comes from tracing over  part of the SYK Hilbert space, there must also be a mechanism to modify the trace that we have yet to understand \cite{Kitaev:2018wpr}, \cite{Blommaert:2018iqz}, \cite{Jafferis:2019wkd}.\footnote{One  puzzling element of this story is that the quantity ``$k \sinh 2\pi k"$ is the Plancherel measure for the semigroup $SL(2,R)^+$ when we restrict to $SL(2,R)$ matrices with positive elements only, and not for $SL(2,R)$ itself \cite{Blommaert:2018iqz}. From the Euclidean perspective, this follows from the fact that the Euclidean Schwarzian field has winding 1 around the thermal circle, imposing a nonlocal constraint in Euclidean time \cite{Jafferis:2019wkd}. See section 4.1 of \cite{Fan:2021wsb} for a demonstration of how restricting to the winding=1 sector changes the measure from the naive Plancherel measure of $SL(2,R)$ to that of $SL(2,R)^+$. The challenge is to understand this from the Lorentzian point of view.} Because of results like \eqref{main2}, \eqref{main22}, \eqref{main222}, this problem is the same as that of understanding the origin of the Bekenstein-Hawking entropy in JT gravity. 

Perhaps these problems would become more tractable if we go up a dimension, which could also put us in a better place to make contact with string theory. In $2+1d$ pure gravity, there is an interesting result resembling \eqref{main2} where the Bekenstein-Hawking entropy of a BTZ black hole is related to a Plancherel measure of a certain noncompact quantum group \cite{McGough:2013gka}. 
If we could more generally relate Ryu-Tayakanagi areas to this Plancherel measure, then since more is known about the role of quantum groups in CFTs than about the $SL(2,R)$ in the SYK model, it could give us new ideas about how to proceed. For example, the modified trace of \cite{Jafferis:2019wkd} may be explicable as a quantum trace. See \cite{Donnelly:2020aa}, \cite{Jiang:2020cqo} for related recent results in the context of Gopakumar-Vafa duality. 

More generally, if we could interpret the RT area term as a $``\log \dim R"$-type edge term for a quantum group symmetry, it would suggest that the analog of the Hilbert space sectors labeled by $SL(2,R)$ representations that were conjectured here for the SYK model should be Hilbert space sectors labeled by conformal families of the CFT associated to that quantum group, in each of the BCFTs that we implicitly split a holographic CFT into when we do an entanglement entropy calculation \cite{Ohmori:2014eia}, \cite{VanRaamsdonk:2018zws}. I will develop this point further in future work.


 
\begin{appendix}
\section{Multiple particle insertions} \label{sa}

In this section, I will generalize the calculation and toy model of section \ref{s3} to the case when we have two operator insertions along the boundary of the Euclidean path integral that sets up the state. This situation was also studied in \cite{Goel:2018ubv}. The generalization to $n > 2$ operator insertions is straightforward.

Namely, we are now interested in the state set up by cutting the Euclidean correlation function
\begin{eqnarray}
Z_1 &=& \left\langle\OO_{1}\left(\tau_1 + \frac {\tau_2}{2} + \tau_3\right)\OO_{1}\left(\frac{\tau_2}{2}+ \tau_3\right) \OO_2(\tau_3)\OO_2(0) \right\rangle \\
&=& \int \prod_{i=1,2,3} dk_i\, k_i\sinh 2\pi k_i \, e^{-\frac{k_1^2}{2C}\tau_1}e^{-\frac{k_2^2}{2C}\tau_2 } e^{-\frac{k_3^2}{2C}\tau_3}\, \frac{\Gamma(\ell_1 \pm ik_1 \pm ik_2)}{\Gamma(2\ell_1)}\frac{\Gamma(\ell_2\pm ik_2 \pm ik_3)}{\Gamma(2\ell_2)}\label{ea2}
\end{eqnarray}
at the moment of time-reflection symmetry,
where $\Delta(\OO_1) = \ell_1$, $\Delta(\OO_2) = \ell_2$, and $\tau_1 + \tau_2 + \tau_3 = \beta$. As \cite{Goel:2018ubv} explained, such states are geometrically described by wormholes with three local horizons, and the entanglement entropy between the two SYK models can be written as 

\be\label{ea3}
S = S_0 + \frac{1}{4G_N} \min (\langle\phi_1\rangle_{\tau_i}, \langle\phi_2\rangle_{\tau_i}, \langle\phi_3\rangle_{\tau_i}) = S_0 + 2\pi \min(k_{1, s.p.}, k_{2, s.p.}, k_{3, s.p.})\,,
\ee
where the $k_{i, s.p.}$'s are the solutions to the saddle-point equations from jointly extremizing the integrand of \eqref{ea2} w.r.t. each of the $k_i$'s. 

\subsection{Rewriting the EE between the two QM's}

	\begin{figure}
	\centering
	\includegraphics[height=1.4in]{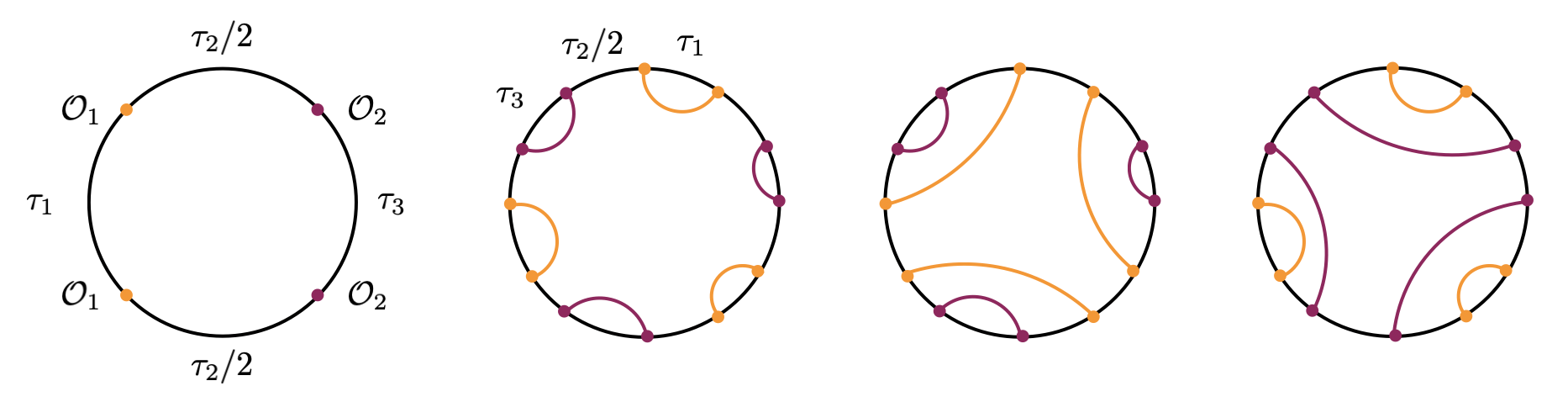}
	\caption{Left: graphical depiction of the Euclidean correlation function \eqref{ea2} that defines the states that we study in this section. Right: The three possibilities for the leading channel that contributes to $Z_n$ are shown here for $n=3$.}
	\label{figa1} 
	\end{figure}

To generalize the steps from section \ref{s3}, we will rewrite \eqref{ea3} in terms of
\be
p_{k_1k_2k_3} = Z_1^{-1}\, \prod_{i=1,2,3}  k_i\sinh 2\pi k_i \, e^{-\frac{k_1^2}{2C}\tau_1}e^{-\frac{k_2^2}{2C}\tau_2 } e^{-\frac{k_3^2}{2C}\tau_3}\, \frac{\Gamma(\ell_1 \pm ik_1 \pm ik_2)}{\Gamma(2\ell_1)}\frac{\Gamma(\ell_2\pm ik_2 \pm ik_3)}{\Gamma(2\ell_2)}\,.
\ee
Note that $Z_1p_{k_1k_2k_3}$ is the integrand of \eqref{ea3}. In this case, there are three possibilities for the leading contribution to the $n$-replicated partition function $Z_n$, 
	in which we connect all of the boundary regions labeled by one of $\tau_1$, $\tau_2$, or $\tau_3$ in the bulk; see Figure \ref{figa1}.
The diagram where we connect all of the regions labeled by $\tau_2$ evaluates to 
(here using the short-hand  $\Gamma_\ell(k, k') = \frac{\Gamma(\ell \pm ik \pm ik')}{\Gamma(2\ell)}$)
\be\begin{split}\label{ea5}
Z_n &\approx \int dk_b\, k_b \sinh 2\pi k_b\, e^{-\frac{n k_b^2}{2C}\tau_2} \int\dots\int \prod_{i=1,\dots,n}dk_{1i}\, k_{1i} \sinh 2\pi k_{1i} \,e^{-\frac{k_{1i}^2}{2C}\tau_1}\Gamma_{\ell_1}(k_{1i}, k_b) \\
&\times \int\dots \int \prod_{j=1, \dots, n} dk_{3j}\, k_{3j} \sinh 2\pi k_{3j}\, e^{-\frac{k_{3j}^2}{2C}\tau_3}\Gamma_{\ell_2}(k_{3j}, k_b)\,,
\end{split}
\ee
while the one where we connect all of the regions labeled by $\tau_1$ evaluates to
\be\begin{split} \label{ea6}
Z_n &\approx \int dk_b\, k_b\sinh 2\pi k_b\, e^{-\frac{nk_b^2}{2C}\tau_1} \\ &\times \int \dots \int  \prod_{i=1, \dots, n} dk_{2i}\, k_{2i} \sinh 2\pi k_{2i}\, e^{-\frac{k_{2i}^2}{2C}\tau_2}\, dk_{3i}\, k_{3i} \sinh 2\pi k_{3i}\, e^{-\frac{k_{3i}^2}{2C}\tau_3}\,\Gamma_{\ell_1}(k_b, k_{2i})\Gamma_{\ell_2}(k_{2i}, k_{3i})\,,
\end{split}
\ee
and the one where we connect all of the regions labeled by $\tau_3$ is similar. 

In terms of $p_{k_1k_2k_3}$, we can rewrite \eqref{ea5} as
\be
Z_n = \int dk_2\, (k_2\sinh 2\pi k_2)^{1-n}\, \left[\iint dk_1\, dk_3 \, Z_1 p_{k_1k_2k_3} \right]^n
\ee
and \eqref{ea6} as 
\be
Z_n = \int dk_1\, (k_1 \sinh 2\pi k_1)^{1-n}\, \left[\iint dk_2\, dk_3\, Z_1p_{k_1k_2k_3} \right]^n\,. 
\ee

Plugging into the replica trick formula \eqref{replica}, we find that the entanglement entropy between the two SYK models at leading order in $1/G_N$ ultimately has the same form as \eqref{main21}, but now with three different possibilities for $i$:
\be\label{ea10}
S = S_0 + \min_{i=1,2,3}\left(\int dk_i\, p_{k_i}(-\log p_{k_i} + \log k_i \sinh 2\pi k_i) \right)\,,
\ee
 where $p_{k_i}$ is now defined by marginalizing over {both} of the other $k_{j \neq i}$'s,
\be
p_{k_i} = \iint \prod_{j \neq i} dk_j\, p_{k_1k_2k_3}\,.
\ee
This equation resembles \eqref{lgtee2} on a lattice with three links.

\subsection{Toy model with three local extrema}

\begin{figure}
	\centering
	\includegraphics[height=1.0in]{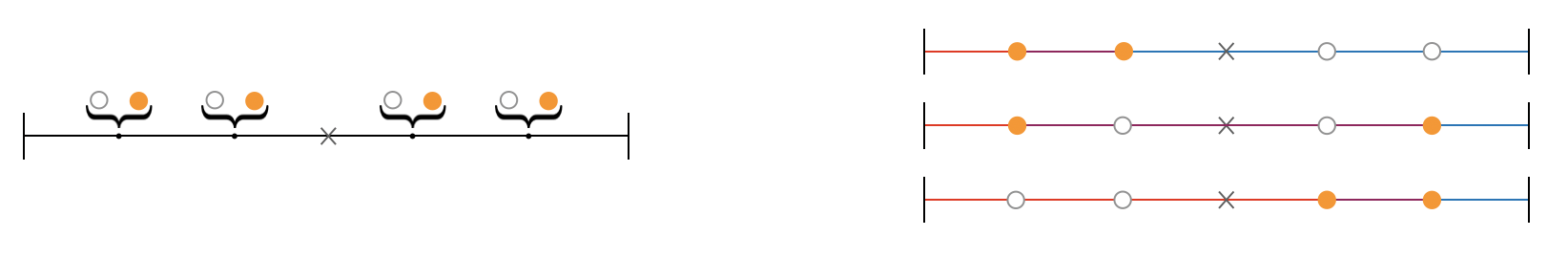}
	\caption{In a compact lattice gauge theory, we can get a result resembling \eqref{ea10} by computing the entanglement entropy with the ``edge mode algorithm" across the location marked "x" in a parametric family of states made from superimposing ones on the right-hand side of this figure. See the caption of Figure \ref{fig4} for the notation that we are using here.}
	\label{figa2} 
	\end{figure}

To construct a situation in a compact lattice gauge theory that allows us to interpolate between three extrema as in \eqref{ea10}, we can consider a lattice with five links and four nodes, and compute the entanglement entropy across the middle link in a superposition of (e.g.) the three states shown in Figure \ref{figa2}. Then everything works the same as in section \ref{s33}. 

Note that although the lattice with five links and four nodes allows for four distinct configurations where the middle link is in the purple representation, we only want to consider superpositions that contain (any) one of them. The point is to mimic a situation where the entanglement cut can go through any one of three flux lines but the setup is otherwise topological. 


\section{Subleading (planar) corrections}\label{sb}

In section \ref{s3}, we rewrote the entanglement entropy between the two SYK models to leading order in $1/G_N$ only. It could be interesting to try to go beyond the leading order and find a formula that interpolates between the minima in \eqref{main21}. We could then use it to make a more precise comparison to  \eqref{e320}. In this section, I will take some steps towards this by setting up resolvent equations that could help us to find the entanglement entropy at planar order. However, I will not manage to find a nice closed-form formula for the entanglement entropy at planar order.

In order to get a subleading result that interpolates between the minima in \eqref{main21}, we need to sum some subset of channels contributing to $Z_n$ that contains both of the leading-order channels and also has a good analytic continuation in $n$. Following \cite{Penington:2019kki}, one possibility is to sum the planar diagrams. 
The claim to fame of the planar diagrams is that a certain combination of them admits a nice recursive definition. Namely, let the {resolvent} $R(k, \lambda)$ be the function s.t. 
\be\label{eb1}
\int dk\, \mu(k) R(k, \lambda) = \lambda^{-1} c + \lambda^{-2} Z_1 + \dots + \lambda^{-(n+1)}Z_n + \dots
\ee
for $\mu(k)= k \sinh 2\pi k$ and $c = \int dk\,\mu(k)$ (note that this is formally infinite). Given $R(k, \lambda)$ in closed form, we can then use it to find the entanglement entropy, see section 2.5 in \cite{Penington:2019kki}. 

For the $Z_n$'s in section \ref{s3}, the Schwinger-Dyson equation for \eqref{eb1} takes the form
\be
\includegraphics[height=1.1in]{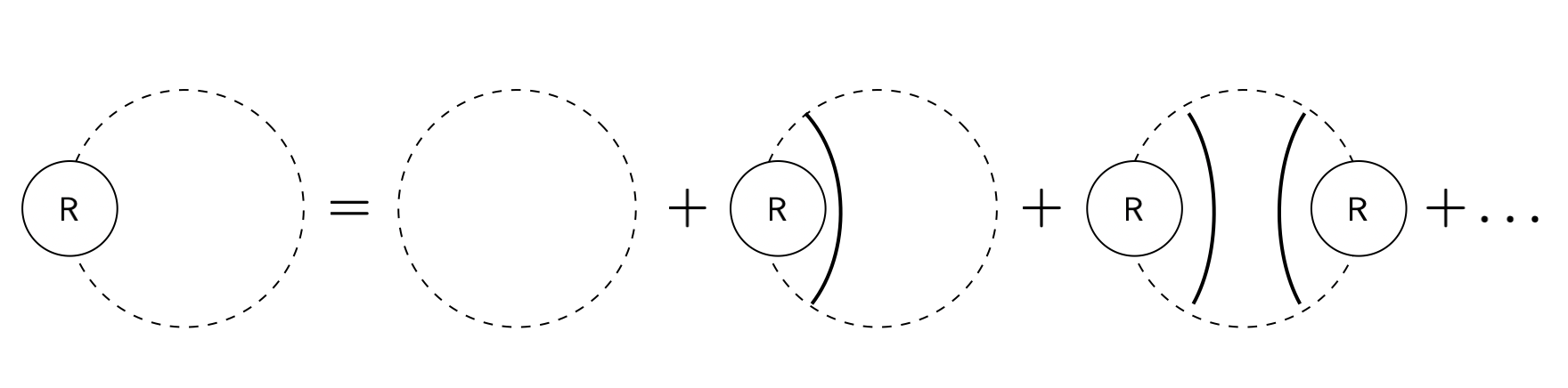}\,,
\ee
or in symbolic form, again using the shorthand $\Gamma_\ell(k, k') = \frac{\Gamma(\ell \pm ik \pm ik')}{\Gamma(2\ell)}$:

\begin{eqnarray}
\int dk\, \mu(k)R(k, \lambda) &=& \sum_{n=0}^\infty \lambda^{-(n+1)}\int dk_b\, \mu(k_b)\, e^{-\frac{n k_b^2}{2C}(\beta - \tau)}\left[\int dk\, \mu(k)e^{-\frac{k^2}{2C}\tau}\Gamma_\ell(k, k_b)R(k,\lambda)\right]^n \\
&=& \sum_{n=0}^\infty \lambda^{-(n+1)} \int dk_b\, \mu(k_b)^{1-n}\left[\int dk\, Z_1p_{kk_b}R(k,\lambda) \right]^n\,.
\end{eqnarray}
As a check, if we plug in $R=1$ on the right-hand side only, we recover the sum over one of the two leading channels from \eqref{e311}. 

Summing over the geometric series on the right-hand side, we find 
\be
\int dk\, \mu(k) R(k, \lambda) = \int dk\, \mu(k) \lambda^{-1} \left(1 - \lambda^{-1}\mu(k)^{-1}\int dk' Z_1 p_{k k'} R(k', \lambda)\right)^{-1}\,,
\ee
or if we assume that we can strip off the integral on both sides, 
\be
\int dk'\, Z_1 p_{kk'} R(k', \lambda) R(k, \lambda) - \lambda \mu(k)\, R(k, \lambda) + \mu(k) = 0\,.
\ee
This is a system of coupled quadratic equations that we can try to solve for $R$. However, I will leave this for future work.


\end{appendix}

\bibliographystyle{ssg}
\bibliography{bhe}		
\end{document}